\documentclass[floats,floatfix,showpacs,amssymb,prd,twocolumn,superscriptaddress,nofootinbib,longbibliography]{revtex4-1}

\usepackage{anyfontsize}
\usepackage{amssymb}
\usepackage{amsmath}
\usepackage{amsfonts}
\usepackage{amsbsy}
\usepackage[utf8]{inputenc} 
\usepackage{txfonts}
\usepackage{mathrsfs}

\usepackage{bm}

\usepackage[utf8]{inputenc}

\usepackage{graphicx}
\usepackage{epsfig}
\usepackage{epstopdf}

\usepackage[usenames,dvipsnames]{xcolor}

\usepackage{verbatim,mathtools,needspace,enumitem,etoolbox}

\definecolor{linkcolor}{rgb}{0.0,0.3,0.5}

\usepackage[unicode, colorlinks=true, linkcolor=linkcolor, citecolor=linkcolor, filecolor=linkcolor,urlcolor=linkcolor, pdfusetitle]{hyperref}

\usepackage{epstopdf}

\usepackage{bm}
\usepackage{dcolumn}

\usepackage{longtable}
\usepackage{enumerate}
 \usepackage[normalem]{ulem}

\hypersetup{colorlinks=true,
            citecolor=blue,
            linkcolor=blue,
            urlcolor=blue}
\usepackage[T1]{fontenc}
\usepackage{subfigure}
\usepackage{amsmath}
\usepackage{float}
\usepackage{mathrsfs}
\usepackage[dvipsnames]{xcolor}
\usepackage{soul}

\definecolor{darkgreen}{RGB}{1,212,57}

\setcitestyle{numbers,square}

\setcitestyle{numbers,square}

\begin{document}
\title{4D Einstein-Gauss-Bonnet gravity: Massless particles and absorption of planar spin-0 waves
}

\author{Haroldo C. D. Lima Junior}
\email{haroldo.ufpa@gmail.com} 
\affiliation{Faculdade de F\'{\i}sica,
Universidade Federal do Par\'a, 66075-110, Bel\'em, PA, Brasil }

\author{Carolina L. Benone}%
 \email{benone@ufpa.br}
\affiliation{%
 Campus Universit{\'a}rio Salin{\'o}polis, Universidade Federal do Par{\'a}, 68721-000, Salin{\'o}polis, Par{\'a}, Brazil 
}%

\author{Lu\'{\i}s C. B. Crispino}
\email{crispino@ufpa.br} 
\affiliation{Faculdade de F\'{\i}sica,
Universidade Federal do Par\'a, 66075-110, Bel\'em, PA, Brasil }

\begin{abstract}
We investigate the absorption cross section of planar 
scalar 
massless waves impinging on spherically symmetric black holes which are solutions of the novel 4D Einstein-Gauss-Bonnet theory of gravity. Besides the mass of the black hole, the solution depends also on the Gauss-Bonnet constant coupling. Using the partial waves approach, we show that the absorption cross section depends on the Gauss-Bonnet coupling constant. Our numerical results present excellent agreement with the low- and high- frequency approximations, including the so-called sinc approximation.
\end{abstract}

\maketitle

\section{Introduction}
General relativity (GR) and quantum field theory (QFT) are the most prominent theories for the description of gravitational interaction and phenomena at the atomic scale, respectively. The agreement of both theories with the experimental data is remarkable. GR has predicted the existence of black holes (BHs), which are supported by the observations on the gravitational wave channel performed by LIGO/VIRGO ~\cite{VL1,VL2,VL3,VL4}, as well as on the electromagnetic wave channel with the first ever image of a BH shadow, captured by the Event Horizon Telescope (EHT) collaboration~\cite{EHT1,EHT2,EHT3,EHT4,EHT5,EHT6}. On the other hand, QFT has predicted, for instance, the value of the fine-structure constant with astonishing accordance with the experimental measurements~\cite{FSC1,FSC2,FSC3}.

Despite the excellent agreement of both GR and QFT with experiments, an unified description of gravity at the Planck scale is still an open problem in theoretical physics. The power-counting   method, introduced by Dyson, indicates that GR is not a renormalizable theory \cite{Wald}. In this context, it is worthwhile to explore alternative theories of gravity which are renormalizable. 

Quadratic-order curvature corrections to GR may lead to renormalizable theories of gravity~\cite{Stelle::1977}. Moreover, such alternative theories of gravity may help to handle the singularity formation problem present in GR~\cite{Hawking::1970}. Among such alternative theories of gravity, there is the Einstein-Gauss-Bonnet (EGB) theory. (See Ref.~\cite{ATG::review} for a review on other theories of gravity with higher curvature corrections.) 

The EGB theory of gravity consists on the Einstein-Hilbert action coupled to the Gauss-Bonnet term~\cite{EGB}. In four dimensions, the Gauss-Bonnet term vanishes identically and does not contribute to the field equations, unless this term is coupled to a matter field, for instance a dilaton field, which is the so called Einstein-dilaton-Gauss-Bonnet theory. The Gauss-Bonnet term contributes to the dynamics of the gravitational field if $D>4$, where $D$ is the number of spacetime dimensions. Notwithstanding, Glavan and Lin have recently claimed that by rescaling the Gauss-Bonnet coupling constant on the EGB theory in $D>4$, and then taking the limit $D\rightarrow 4$, it is possible to obtain nontrivial contributions from the Gauss-Bonnet term in four dimensions~\cite{4D_EGB}, and to bypass the Lovelock's theorem~\cite{Lovelock::1971}. 
Moreover, in Ref.~\cite{4D_EGB}, a BH solution of this 4D-EGB theory was obtained, which reduces to the Schwarzschild solution in the limit of vanishing Gauss-Bonnet coupling constant.

However, a number of authors pointed out that this dimensional limit taking method is ill-defined~\cite{Gurses::2020arxiv, Arrechea:2020arxiv, Hennigar:2020arxiv}. For instance, divergent contributions to the field equations in four dimensions were observed~\cite{Arrechea:2020arxiv}, as well as the lack of covariant description~\cite{Gurses::2020arxiv}. A consistent formulation of the $D\rightarrow 4$ limit in EGB theory and the subtleties of this method were discussed in Ref.~\cite{Aoki:2020arxiv} using the ADM decomposition, and it was pointed out that the theory either breaks the diffeomorphism invariance or has additional degrees of freedom, thus in accordance with the Lovelock's theorem~\cite{Lovelock::1971}. Nevertheless, the 4D-EGB theory can also be obtained as a particular case of the Horndeski theory~\cite{Hennigar:2020arxiv}. In any case, the original solution proposed in~\cite{4D_EGB} is also solution for the consistent formulations of the 4D-EGB theory~\cite{Hennigar:2020arxiv,Aoki:2020arxiv}.

The interaction of BHs with fields may lead to interesting phenomena, for instance, absorption~\cite{Abs1,Abs2,Abs3,Abs4,Abs5,Abs6,Abs7,Abs8,Abs9,Abs10,Abs11}, scattering~\cite{SC:2006,SC:2014,SC:2015,SC:2015.1}, and radiation emission~\cite{R_EM, R_EM2, R_EM3, R_EM4}. In particular, the absorption and scattering of fields are of major interest in the description of fields around BHs, since they are related to processes occurring in active galactic nuclei (AGN)~\cite{AGN1,AGN2}. Spin-0 fields model scalar particles, or processes where the spin does not play a major role. Furthermore, spin-0 fields may also be associated to some dark matter candidates~\cite{Boehm::2004}.

In this letter, we study for the first time the absorption of planar massless spin-0 waves in 4D-EGB gravity, using the partial wave approach. 
In particular, we investigate the effects of the EGB coupling constant in the absorption cross section. We compare our results to the spherically symmetric vacuum solution of GR, i.e., the Schwarzschild solution, and find that they may be quite distinctive. In order to compare our numerical results with the high-frequency approximation, we also study some features of null geodesics in the EGB theory, for instance, the light rings (LRs) and the geometric capture cross section, which is related to the shadow of the BH.

The remainder of this letter is organized as follows. In Sec.~\ref{Spacetime}, we present the BH solution of the 4D-EGB theory of gravity, and some of its properties. Section~\ref{ABS} is dedicated to different aspects of the absorption cross section. In Subsec.~\ref{CPS}, we study the null geodesics in the 4D-EGB BH spacetime, and find the total capture cross section. Subsections~\ref{Sinc}  and \ref{lfl} are dedicated to the sinc and low-frequency approximations, respectively. In Subsec.~\ref{PWA}, we present our numerical results, obtained with the partial wave approach. Our final remarks are presented in Sec.~\ref{Final_remarks}. Throughout this letter, we set $G=c=\hbar=1$.

\section{The spacetime}
\label{Spacetime}
As a spherically symmetric solution in the 4D-EGB gravity, we can write~\footnote{In comparison with Refs.~\cite{EGB,4D_EGB}, 
the Gauss-Bonnet coupling constant exhibited in our Eq.~\eqref{line_el} differs by a multiplicative factor of $16\pi$.
This choice simplifies the notation and does not affect the results.} \cite{4D_EGB}
\begin{equation}
\label{line_el}ds^2=-f(r)\,dt^2+\frac{1}{f(r)}\,dr^2+r^2\,d\Omega^2,
\end{equation}
where $d\Omega^2$ is the line element of the unit sphere and
\begin{equation}
\label{f_r}f(r)\equiv 1+\frac{r^2}{2\,\alpha}\left[1\pm \left(1+\frac{8\,\alpha\,M}{r^3}\right)^{1/2} \right],
\end{equation}
with $M$ being the ADM mass of the BH. In Eq.~\eqref{f_r}, we choose the minus sign, otherwise the spacetime is not asymptotically flat. The horizons of the BH described by line element \eqref{line_el} are given by
\begin{equation}
r_{\pm}\equiv M\left[1\pm \sqrt{1-\frac{\alpha}{M^2}}\right],
\end{equation}  
where $r_+$ is the radial location of the BH event horizon. We note that the event horizon radial coordinate is real valued only for $\alpha<M^2$. Moreover, the spacetime is well behaved outside the event horizon also for negative values of $\alpha$, such that 
\begin{equation}
\label{alpha_domain}-8M^2\leq\alpha\leq M^2.
\end{equation} 
It was shown that 4D-EGB BHs with large coupling constant $\alpha$ are unstable under gravitational pertubations~\cite{4D_EGB19}. Additionally, for the negative branch of $\alpha$, the line element is  not well defined for values of the radial coordinate $r<r_0$, where
\begin{equation}
1+\frac{8\alpha M}{r_0^3}=0.
\end{equation}
Our results are valid for positive as well as negative values of $\alpha$.
As we take the limit $\alpha\rightarrow 0$, we recover the well known Schwarzschild spacetime, which is the vacuum solution of GR. Many properties of the spacetime \eqref{line_el} have been studied recently in a number of papers, for instance, in 
Refs.~\cite{4D_EGB1, 4D_EGB2,4D_EGB3,4D_EGB4,4D_EGB5,4D_EGB6,4D_EGB7,4D_EGB8,4D_EGB9,4D_EGB12,4D_EGB13,4D_EGB14,4D_EGB15,4D_EGB16,4D_EGB17,
4D_EGB18}.

\section{Absorption cross section}
\label{ABS}

\subsection{Null geodesics and capture cross section}
\label{CPS}
In the high-frequency regime, massless spin-0 waves are effectively described by null geodesics. Null geodesics, gravitational lensing and shadows in the 4D-EGB BH were recently studied in Refs.~\cite{4D_EGB1,4D_EGB2,4D_EGB14}. Since the line element \eqref{line_el} is spherically symmetric, we may restrict our analysis to the equatorial plane without loss of generality. Hence, the motion of massless particles can be described by:
\begin{equation}
\label{eq_motion1}-f(r)\dot{t}^2+\frac{1}{f(r)}\dot{r}^2+r^2\dot{\phi}^2=0,
\end{equation}
where the dots represent differentiation with respect to the affine parameter $\tau$ along the geodesics. Due to the existence of the Killing vectors
\begin{align}
&k^{\mu}=\left(\frac{\partial}{\partial t}\right)^\mu,\\
&\omega^{\mu}=\left(\frac{\partial}{\partial \phi}\right)^\mu,
\end{align}
in $t$ and $\phi$ direction, respectively, we have two conserved quantities
\begin{align}
\label{tdot}&\dot{t}=\frac{E}{f(r)},\\
\label{phidot}&\dot{\phi}=\frac{L}{r^2},
\end{align} 
which are related to the energy ($E$) and angular momentum ($L$) of the massless particle. Inserting Eqs.~\eqref{tdot} and \eqref{phidot} into Eq.~\eqref{eq_motion1}, we obtain an \textit{energy balance} equation:
\begin{equation}
\label{E_balance}\dot{r}^2+V_{eff}=E^2,
\end{equation}
where $V_{eff}$ is the effective potential for massless particles in 4D-EGB BH spacetime, defined as
\begin{align}
\label{V_eff}V_{eff}\equiv L^2\left[\frac{1}{r^2}+\frac{1}{2\alpha}\left(1-\sqrt{1+\frac{8\alpha M}{r^3}}\right)\right].
\end{align}

Around BHs, light can be bent in arbitrarily large angles, and even describe planar closed orbits with constant radial coordinate. Such planar closed orbits, dubbed as light rings (LRs), are generically present in BH configurations~\cite{Cunha::2020}. For static and spherically symmetric spacetimes, they satisfy $\dot{r}=\ddot{r}=0$. Using Eq.~\eqref{E_balance}, we write the equations of a LR as
\begin{align}
\label{SPO1}&V_{eff}=E^2,\\
\label{SPO2}&\frac{dV_{eff}}{dr}=0.
\end{align}
Rewritting Eqs.~\eqref{SPO1} and \eqref{SPO2} in terms of the impact parameter, $b \equiv L/E$, and solving the corresponding equations, we find the radius of the LR, $r_p$, and the critical impact parameter, $b_c$, to be
\begin{align}
\label{radius_SPO}&r_p=\frac{3M^2+\beta^{2/3}}{\beta^{1/3}},\\
\label{bc_SPO}&b_c=\frac{r_p}{\sqrt{f(r_p)}}.
\end{align}
respectively. In Eq.~\eqref{radius_SPO}, $\beta$ is given by
\begin{equation}
\beta\equiv -4M\alpha + \sqrt{16M^2\alpha^2-27M^6}.
\end{equation}
The critical impact parameter $b_c$ determines the threshold between scattered and absorbed null geodesics. Massless particles with $b>b_c$ are scattered by the BH, while massless particles with $b<b_c$ are absorbed by the BH. 
Since the spacetime is spherically symmetric, $b_c$ is the radius of the shadow of the 4D-EGB BH, as seen by a distant observer. 
We note that the critical impact parameter $b_c$ is a monotonically decreasing function of the Gauss-Bonnet coupling parameter. 
In the high-frequency regime, the total absorption cross section of the black hole is the geometric capture cross section, given by 
\begin{equation}
\label{sigma_geo}\sigma_{geo}\equiv\pi\,b_c^2.
\end{equation}

\subsection{Sinc approximation}
\label{Sinc}
We can improve the high-frequency analysis, described by null geodesics, by considering the so-called sinc approximation~\cite{Abs1,Sinc_approx}. In the sinc approximation, the total absorption cross section is written as a sum of the geometric capture cross section plus an oscillatory part \cite{Sinc_approx}, i.e.
\begin{align}
\sigma\approx \sigma_{geo}+\sigma_{sinc},
\end{align}
where
\begin{equation}
\sigma_{sinc}\equiv-\frac{8\pi\lambda_p}{\Omega_p}e^{\pi\lambda_p/\Omega_p}\,\text{sinc}\left(\frac{2\pi\omega}{\Omega_p}\right)\,\sigma_{geo} ,
\end{equation}
\begin{equation}
\text{sinc}\hspace{0.02in} (x) \equiv \frac{\sin x}{x},
\end{equation}
\begin{equation}
\lambda_p\equiv\left.\frac{1}{\dot{t}}\sqrt{\frac{1}{2}\frac{d^2}{dr^2}\dot{r}^2}\right|_{r=r_p}
\end{equation}
is the Lyapunov exponent and 
\begin{equation}
\Omega_p\equiv\left.\frac{d\phi}{dt}=\sqrt{\frac{f(r)}{r^2}}\right|_{r=r_p}
\end{equation}
is the angular coordinate velocity computed at the LR radius. 
In Subsec.~\ref{PWA} we compare our numerical results with the ones obtained {\it via} 
the sinc approximation.

\subsection{Low-frequency limit}
\label{lfl}
It is well known that the total absorption cross section of massless spin-0 waves tends to the area of the BH event horizon in the low-frequency regime~\cite{Das_Gibbons_Mathur}. This result is valid for static, as well as stationary BHs~\cite{AH1}. 
The general results presented in Refs.~\cite{Das_Gibbons_Mathur,AH1} do not depend on the explicit form of the scalar field equations, and they remain valid for the 4D-EGB theory. This is confirmed with the numerical results presented in Sec.~\ref{PWA}, where we note that the total absorption cross section per unit area of the BH event horizon tends to the unity, as the frequency goes to zero. Moreover, we note that the area of the event horizon is a monotonically decreasing function of the Gauss-Bonnet parameter in the domain given by Eq. \eqref{alpha_domain}.

\begin{center}
\begin{figure}[ht!]
\center
\includegraphics[scale=0.55]{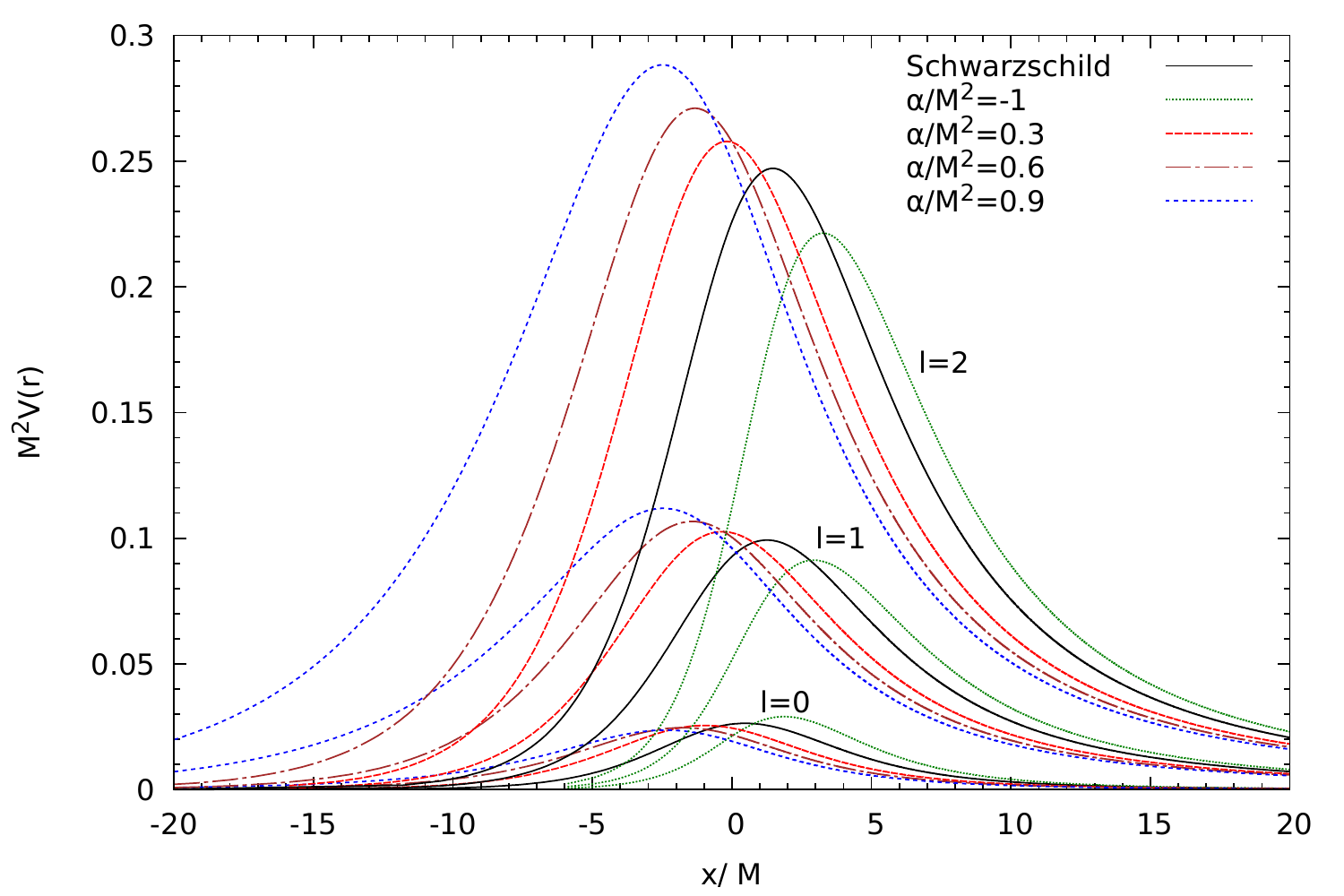}
\caption{Scattering potential of the spin-0 field, given by Eq.~(\ref{V}), for different values of the Gauss-Bonnet coupling $\alpha$. We also show the effective potential for the Schwarzschild spacetime ($\alpha=0$), for comparison.}
\label{V_scalar_field}
\end{figure}
\end{center}

\subsection{Partial wave approach}
\label{PWA}
We now study the absorption cross section of massless spin-0 waves using the partial wave approach method. The dynamical evolution of the field $\Phi$ is described by the Klein-Gordon equation, namely
\begin{equation}
\label{field_eq}\frac{1}{\sqrt{-g}}\partial_\mu\left(\sqrt{-g}g^{\mu\nu}\partial_\nu\Phi\right)=0,
\end{equation}
where g is the metric determinant, and $g^{\mu\nu}$ are the contravariant components of the metric tensor. We write the field in a spherical harmonics basis [$Y_{lm}(\theta,\phi)$] as follows:
\begin{equation}
\label{sphe_harm_dec}\Phi=\sum_{l,m}\frac{\phi_{lm}(t,r)}{r}Y_{lm}(\theta,\phi).
\end{equation}
Substituting Eq.~\eqref{sphe_harm_dec} into Eq.~\eqref{field_eq}, we obtain
\begin{equation}\label{field_tr}
\begin{split}
&f(r)\frac{\partial}{\partial r}\left(f(r)\frac{\partial \phi_{lm}(t,r)}{\partial r}\right)-\left(\frac{f'(r)}{r}+\frac{l(l+1)}{r}\right)\phi_{lm}(t,r)\\
&-\frac{\partial^2\phi_{lm}(t,r)}{\partial t^2}=0,
\end{split}
\end{equation}
where $f'(r)\equiv df(r)/dr$. Applying an inverse Fourier transformation to $\phi_{lm}(t,r)$, we obtain
\begin{equation}
\label{fourier_transf}\phi_{lm}(t,r)=\frac{1}{\sqrt{2\pi}}\int^{+\infty}_{-\infty}e^{-i\omega t}\phi_{lm}(\omega,r)d\omega,
\end{equation}
and substituting Eq.~\eqref{fourier_transf} into Eq.~\eqref{field_tr}, we are left with
\begin{align}
\label{Radial_eq}\frac{d^2\phi_{lm}(\omega, x)}{dx^2}+\left(\omega^2-V\right)\phi_{lm}(\omega, x)=0,
\end{align}
where $dx\equiv\frac{1}{f(r)}dr$ is the tortoise coordinate in the 4D-EGB spacetime, and $V(r)$ is the scattering potential for the scalar field, given by
\begin{equation}
V(r)=f(r)\left[\frac{f'(r)}{r}+\frac{l(l+1)}{r^2}\right].
\label{V}
\end{equation}

In Fig.~\ref{V_scalar_field}, we plot the scattering potential in terms of the tortoise coordinate for different values of the Gauss-Bonnet coupling and for different values of $l$ and $\alpha$. We note that the peaks of the scattering potential increase (decrease) as we increase the values of $\alpha$ for $l\neq 0$ ($l=0$). Moreover, the potential is zero at the event horizon ($x=-\infty$) and at the spatial infinity ($x=+\infty$). Hence, the appropriate boundary conditions to the scattering/absorption problem are given in terms of the so-called \textit{in modes}:
\begin{equation}
\label{BC}\phi_{lm}(r) \approx \begin{cases} e^{-i\omega x}+\mathcal{R}_{\omega l}e^{i\omega x}, & x\rightarrow +\infty,\\
\mathcal{T}_{\omega l}e^{-i\omega x}, &  x \rightarrow -\infty,
\end{cases}
\end{equation}
where 
$\left|\mathcal{T}_{\omega l}\right|^2$ and $\left|\mathcal{R}_{\omega l}\right|^2$ are the transmission and reflection coefficients, respectively, obeying
\begin{align}
\left|\mathcal{T}_{\omega l}\right|^2+\left|\mathcal{R}_{\omega l}\right|^2=1.
\end{align}
These {\it in modes} were used to compute the greybody factors and Hawking radiation of the 4D-EGB BH in Ref.~\cite{4D_EGB17} and of the 4D-EGB de-Sitter solution in Ref.~\cite{4D_EGB18}.

The partial absorption cross section for spin-0 fields can be written in terms of the transmission coefficient as~\cite{Futterman}:
\begin{equation}
\sigma_l=\frac{\pi}{\omega^2}\left(2l+1\right)\left|\mathcal{T}_{\omega l}\right|^2,
\end{equation}
and the total absorption cross section is given by
\begin{equation}
\sigma_{abs}=\sum_{l=0}^{\infty}\sigma_l.
\end{equation}
We compute numerically the transmission and reflection coefficients by integrating Eq.~\eqref{Radial_eq}, from the event horizon towards spatial infinity, using the boundary conditions \eqref{BC}. 

\begin{center}
\begin{figure}[ht!]
\center
\includegraphics[scale=0.55]{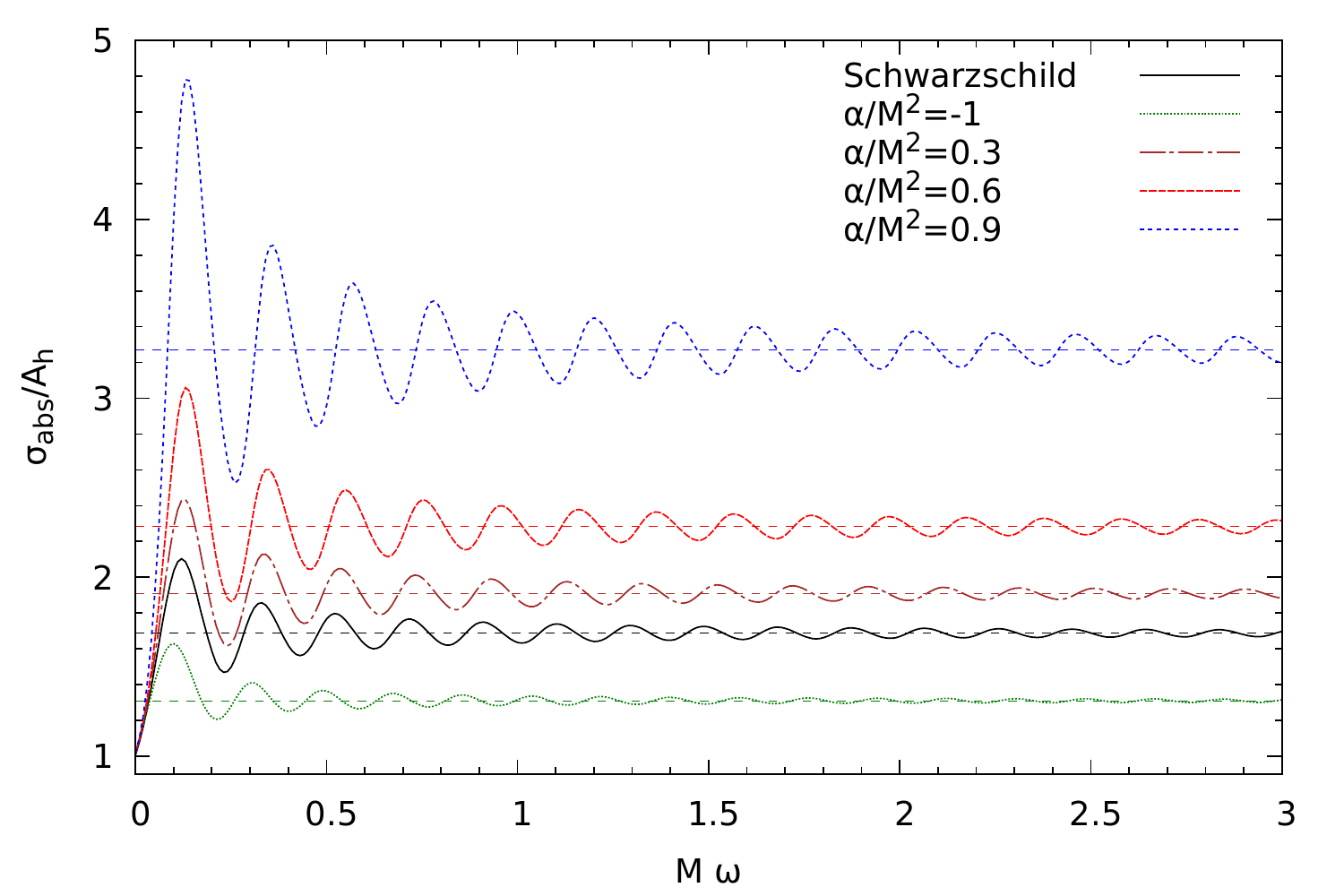}
\caption{Total absorption cross section of planar massless spin-0 waves for spherically symmetric BHs in 4D-EGB theory with different values of Gauss-Bonnet coupling. 
The corresponding geometric (capture) cross sections are plotted as horizontal dashed lines. 
We also plot the total absorption cross section for the Schwarzschild BH ($\alpha=0$).}
\label{Tot_abs}
\end{figure}
\end{center}

\begin{center}
\begin{figure}[ht!]
\center
\includegraphics[scale=0.66]{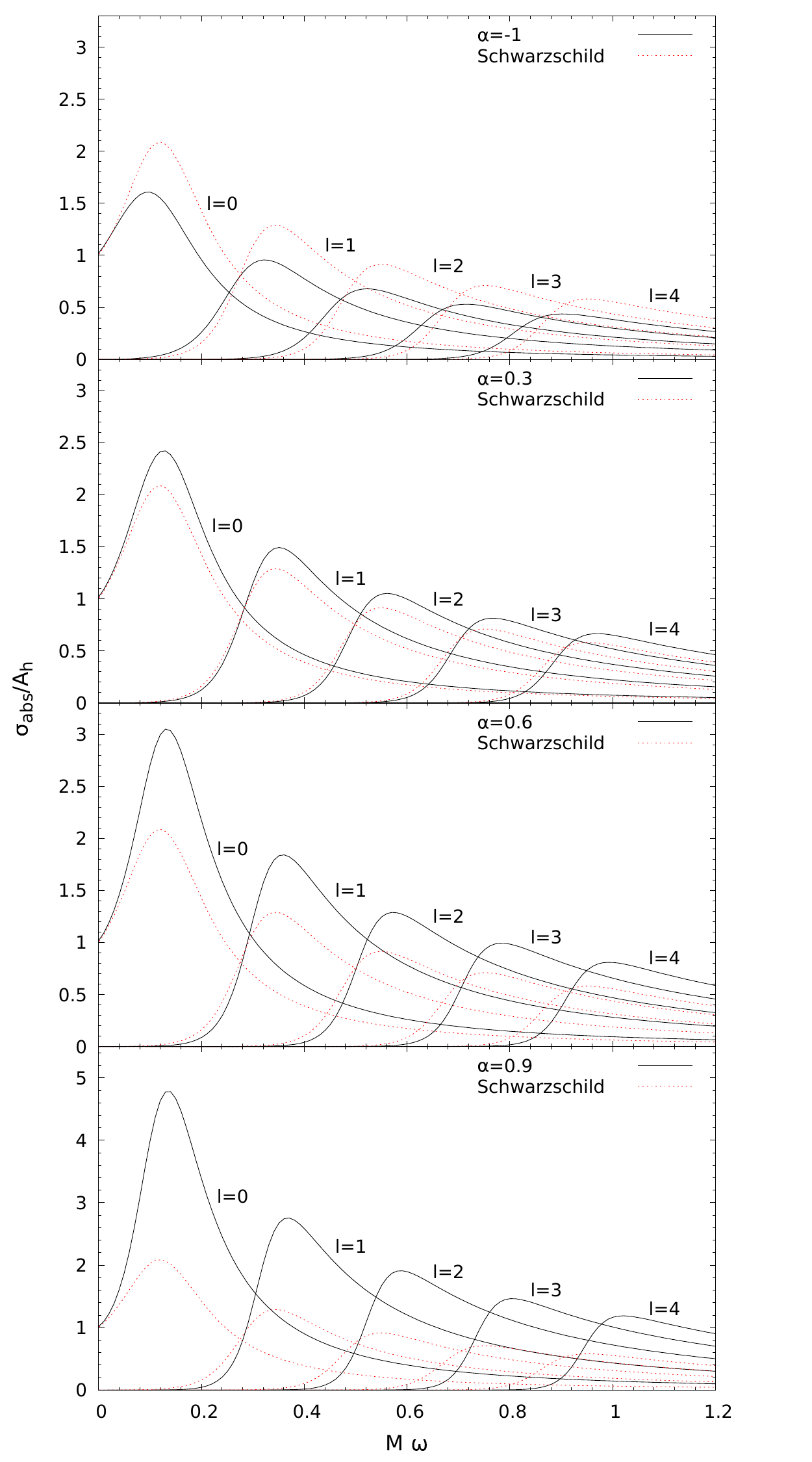}
\caption{Partial absorption cross sections of planar massless spin-0 waves for spherically symmetric BHs in 4D-EGB theory, with different values of Gauss-Bonnet coupling, compared with the Schwarzschild results.}\label{Partial_Abs}
\end{figure}
\end{center}

The results for the total absorption cross section are presented in Fig.~\ref{Tot_abs}, where we have chosen different values for the Gauss-Bonnet coupling $\alpha$.  
The total absorption cross section for the Schwarzschild is also presented in Fig.~\ref{Tot_abs}, for comparison purposes. The regular oscillatory pattern (common in spherically symmetric BH spacetimes) is manifest in the 4D-EGB BH solution. Moreover, as we increase the Gauss-Bonnet coupling constant, the total absorption cross section per unit of event horizon area increases. 
The horizontal dashed lines represent the geometric capture cross section, given by Eq.~\eqref{sigma_geo}. We note in Fig.~\ref{Tot_abs} that the total absorption cross section tends to the capture cross section in the high-frequency regime, as anticipated in Subsec.~\ref{CPS}.

In Fig.~\ref{Partial_Abs}, we show the partial absorption cross section for different values of $l$. We note that the $l=0$ mode is dominant in the low-frequency regime. Besides that, $\sigma_0$ tends to the area of the BH event horizon, as anticipated in Subsec.~\ref{lfl}. This behavior is also manifest in the total absorption cross section, as shown in Fig.~\ref{Tot_abs}. From Figs.~\ref{Tot_abs} and~\ref{Partial_Abs}, we note that the absorption cross section for vacuum BH solutions in GR and 4D-EGB theories of gravity can be quite distinctive, depending on the value of the Gauss-Bonnet coupling constant.

\begin{center}
\begin{figure}[ht!]
\center
\includegraphics[scale=0.55]{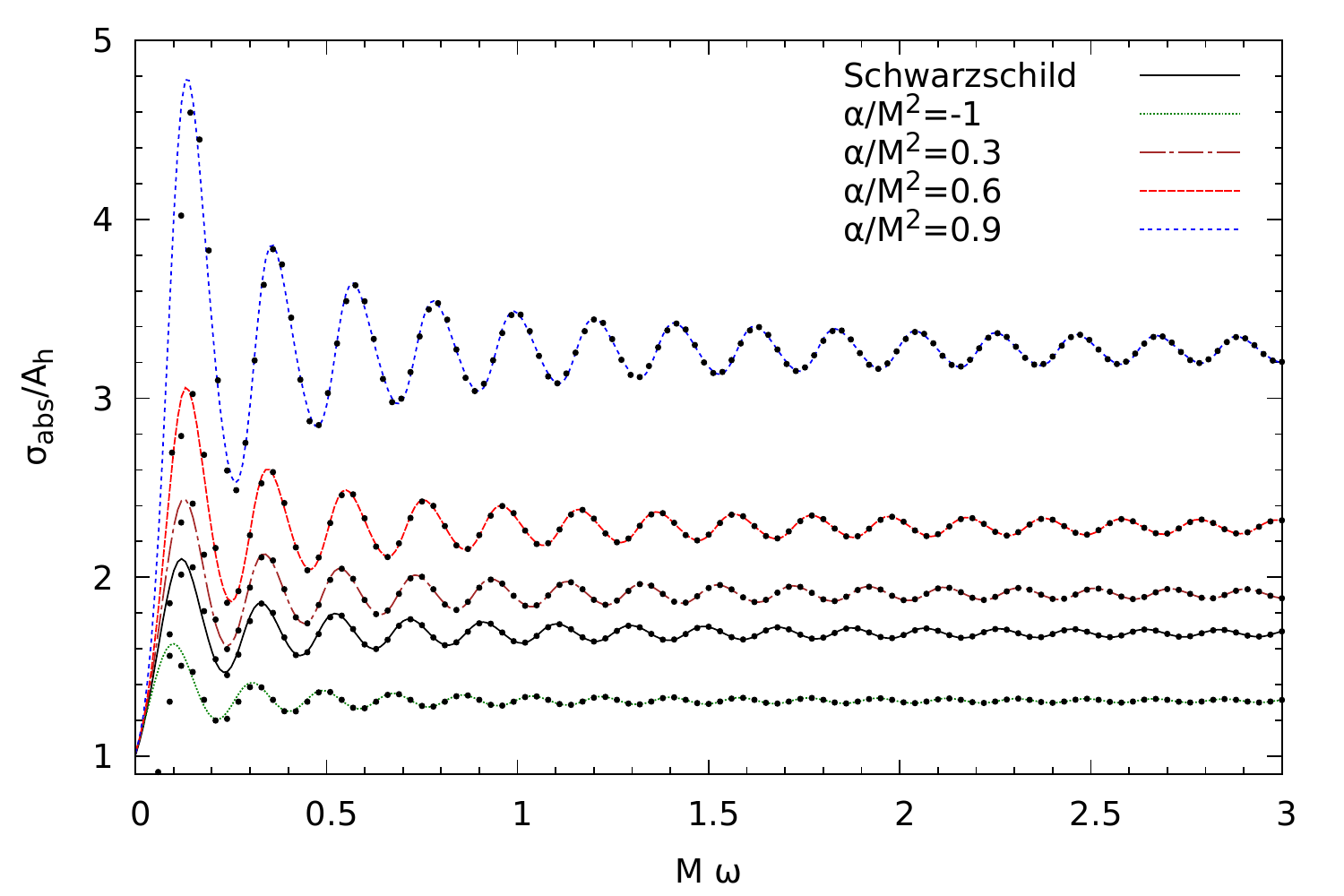}
\caption{Numerical results for the total absorption cross section of spin-0 waves in spherically symmetric BHs in 4D-EGB theory compared with the sinc approximation. The continuous lines corresponds to the numerical results, while the black dots correspond to the sinc approximation.}
\label{Tot_abs_sinc}
\end{figure}
\end{center}

In Fig.~\ref{Tot_abs_sinc} we present the full numerical results compared to the sinc approximation. We note that the results obtained with the two different methods agree remarkably well in the mid-to-high frequency regime.   

\section{Final remarks}
\label{Final_remarks}
We analyzed the absorption of spin-0 massless waves in a spherically symmetric BH spacetime, in the context of the novel 4D-EGB theory of gravity. We have investigated how the  absorption cross section varies with the Gauss-Bonnet coupling constant showing that, for instance, for positive (negative) values of the coupling constant $\alpha$, the total absorption cross section per unit of event horizon area increases (decreases), except at the low-frequency regime. 

Our numerical results are in excellent agreement with the low- and high- frequency approximations. In the low-frequency regime, the total absorption cross section tends to the area of the BH event horizon. On the other hand, for high frequencies the total absorption cross section tends to the (geometric) capture cross section. Moreover, our numerical results also present remarkable agreement with the sinc approximation in the high, and even intermediate, frequencies. The results of the total absorption cross section in the  high-frequency regime are related to the radius of the BH shadow which is of astrophysical importance, specially in the context of the recent results of the EHT collaboration.

\section*{Acknowledgments}
The authors thank 
Conselho Nacional de Desenvolvimento Cient\'ifico e Tecnol\'ogico (CNPq) and Coordena\c{c}\~ao de Aperfei\c{c}oamento de Pessoal de N\'{\i}vel Superior (Capes) - Finance Code 001, for partial financial support.
This research has also received funding from the European Union's Horizon 2020 research and innovation programme under the H2020-MSCA-RISE-2017 Grant No. FunFiCO-777740.


\end{document}